\documentclass{JHEP3}

\usepackage{graphicx}
\usepackage{amssymb}
\usepackage[mathscr]{eucal}
\usepackage{amsmath}
\usepackage{cite}
\usepackage{afterpage}

\newcommand{\gsimm}{\raise.3ex\hbox{$>$\kern-.75em\lower1ex\hbox{$\sim$}}}
\newcommand{\lsimm}{\raise.3ex\hbox{$<$\kern-.75em\lower1ex\hbox{$\sim$}}}

\newcommand{\be}{\begin{equation}}
\newcommand{\ee}{\end{equation}}
\newcommand{\ba}{\begin{eqnarray}}
\newcommand{\ea}{\end{eqnarray}}

\newcommand{\bea}{\begin{eqnarray*}}
\newcommand{\eea}{\end{eqnarray*}}

\def\tr{{\rm{tr}}}

\title{Conformal Inflation Coupled to Matter}

\author{Philippe Brax \\
  Institut de Physique Th\'eorique, CEA, IPhT, CNRS, URA 2306,
  F-91191Gif/Yvette Cedex, France \\ E-mail:
  \email{philippe.brax@cea.fr}}

\author{Anne-Christine Davis\\
  DAMTP, Centre for Mathematical Sciences, University of Cambridge,
  CB3 0WA, UK\\E-mail:
  \email{A.C.Davis@damtp.cam.ac.uk}}
\date{today}

\abstract{We formulate new conformal models of inflation and dark energy which generalise the Higgs-Dilaton scenario. We embed these models in unimodular gravity whose effect is to break scale invariance in the late time Universe. In the early Universe, inflation occurs close to a maximum of both the scalar potential and the scalar coupling to the Ricci scalar in the Jordan frame. At late times, the dilaton, which decouples from the dynamics during inflation, receives a potential term from unimodular gravity and leads to the acceleration of the Universe.   We address two central  issues in this scenario. First we show that the Damour-Polyalov mechanism, when non-relativistic matter is present prior to the
start of inflation, sets the initial conditions for inflation at the maximum of the scalar potential. We then show that conformal invariance implies that matter particles are not coupled to the dilaton in the late Universe at the classical level. When fermions acquire masses at low energy, scale invariance is broken and quantum corrections induce a coupling between the dilaton and matter which is still small enough to evade the gravitational constraints in the solar system. }

\begin{document}


\section{Introduction}
\label{sec:introduction}

The absence of any direct evidence in favour of  physics beyond the standard model at the Large Hadron Collider (LHC), and the discovery that the Higgs mass around 125 GeV could render the standard model viable up to extremely large
energy without the need for any additional features \cite{Buttazzo:2013uya}, may be a hint that both inflation \cite{Martin:2013tda} and the late time acceleration of the Universe \cite{Copeland:2006wr} could have an explanation within the standard model too. Recently, the Higgs field has been considered as a potential candidate for the inflaton when coupled to gravity in a non-trivial way\cite{Bezrukov:2007ep,Bezrukov:2010jz}. By doing this, the resulting spectral index of cosmological perturbations has been found to be close to $0.97$ well within the ballpark \cite{Martin:2013nzq} of the recent observations by the Planck satellite \cite{Ade:2013uln}.

The fact that inflation must end forces the energy density of the inflaton field to vanish at the end of inflation when the Higgs field eventually settles at its minimum. This precludes the existence of a small cosmological constant at the end of inflation as reheating would never take place. This is particularly relevant as the energy density in the symmetric Higgs vacuum is 60 orders of magnitudes larger than the vacuum energy density of the present Universe. This phenomenological fact implies that a large fine-tuning of the vacuum energy at the end of inflation must be imposed in this scenario. This can be avoided if the scale invariance of the Higgs model for large values of the Higgs field, with scale invariance being broken only by the negative mass squared term in the Higgs potential, is  promoted to a full invariance of the model in the absence of gravity\cite{GarciaBellido:2011de}. This requires the existence of one extra field, the dilaton, associated with the Goldstone direction where  the minimum of  the classical potential is flat with a vanishing energy density. Threshold effects, every time a massive particle decouples, induce large corrections to this vacuum energy and jeopardise the potential compatibility with the small vacuum energy inferred from the observations of the acceleration of the Universe. One proposal advocated by Weinberg \cite{Weinberg:1988cp} and originally proposed by Einstein considers the determinant of the space-time metric to be constant. In this unimodular gravity, quantum corrections due to matter do not contribute to the gravitating part of the vacuum energy. This provides a solution to the cosmological constant problem in as much as these quantum corrections to the vacuum energy do not gravitate. However, the fact that the determinant of the metric is fixed implies that scale invariance is broken and a new scale, which can be seen as a Lagrange multiplier, is generated and depends only on the initial conditions in the very early Universe. This constant plays the role of a gravitating cosmological constant. A similar approach to the dark energy problem has been followed recently in \cite{Kaloper:2013zca} where radiative corrections in the matter sector are sequestered. In the context of unimodular gravity and  the Higgs-Dilaton models\cite{GarciaBellido:2011de}, the dilatonic Goldstone direction is lifted by the gravitating cosmological constant term and, in the Einstein frame, it leads to a runaway potential along which the dilaton can play the role of dark energy.

In Higgs inflation models, the initial condition for the Higgs field must be set to be  very large compared to the electroweak scale. The naturalness of this initial condition cannot be assessed within the original proposal. On the other hand, in the Higgs-Dilaton models and in the presence of non-relativistic matter before the start of inflation, we find that there is a natural symmetron mechanism \cite{Hinterbichler:2010es} whereby the normalised inflaton field is forced to be stuck at the origin of field space without any tuning. When the matter density has been redshifted enough and the curvature of the effective potential for the inflaton has become negative, the inflaton field begins  its slow roll evolution and inflation starts\cite{Dong:2013swa}. Then the  non-relativistic matter density decreases extremely rapidly leaving only the inflaton as the main source of energy-density.
The necessary non-relativistic matter density for this mechanism to work could be naturally provided by very heavy particles arising in grand unified theories or as very heavy states in string theory.

At late time the dilaton, which rolls to large values and whose energy density leads to the late time acceleration of the expansion of the Universe, couples to matter and acts as a potential fifth force. No screening mechanism is available for this field and therefore gravity tests in the solar system can only be evaded if the coupling strength of the dilaton to matter is small enough \cite{Bertotti:2003rm}. In fact, we find that the dilaton field decouples from matter at the classical level thanks to the original conformal invariance of the model. At the quantum level, graviton loops induce a coupling between the dilaton and matter when fermions become massive. The gravity tests in the solar system such as the Cassini experiment are satisfied, and the dilaton can play the role of dark energy, when this coupling is small. We find that this is always the case when the gauge boson masses are much lower than  the Planck scale.

We show in this paper that the features that we have just presented about the Higgs-Dilaton model, and the new mechanisms leading to the proper initial conditions for inflation and the absence of fifth force  at low energy, are always present in a much larger class of conformal models which we describe here. We find that inflation  occurs under generic conditions when the scalar potential and the coupling to the Ricci scalar in the Jordan frame have a common maximum and analyse how the Damour-Polyakov \cite{Damour:1994zq} mechanism, which generalises the symmetron mechanism,  sets the initial conditions for inflation. The absence of coupling to matter for the dilaton at the classical level is  still guaranteed and gravitational tests of fifth forces are still evaded when gauge boson masses are much lower than the Planck scale.

Conformal and superconformal models have been  extensively studied recently in relation with Starobinski's model of inflation and its generalisations \cite{Kallosh:2013oma,Kallosh:2013yoa,Kallosh:2013tua,Kallosh:2013daa,Kallosh:2013lkr,Kallosh:2013pby,Kallosh:2013hoa}. Other supersymmetric extensions are also particularly promising\cite{Buchmuller:2013zfa,Ellis:2013nxa,Ellis:2013xoa}. The analysis of the Damour-Polyakov mechanism for all these models is left for future work.

The present paper is arranged as follows. In the first part, we present the Damour-Polyakov mechanism and its application to the early inflationary dynamics. Then in section III, we define the conformal models of inflation and describe some of their features including how the initial conditions are set by the Damour-Polyakov mechanism. We  analyse the late time properties of the models in section IV and the coupling of the dilaton to matter.  In section V, we revisit the Higgs-Dilaton models in view of our general construction. We use the Higgs-Dilaton model as an explicit example of our mechanisms, both showing how the initial conditions of inflation are set by the coupling to non-relativistic matter and how the dilaton field becomes the dark energy particle at late times.
Finally we conclude in section VI.

\section{The Early Damour-Polyakov Effect}

We first review the Damour-Polyakov
effect in scalar-tensor gravity \cite{Damour:1994zq}, which we apply to inflationary models
to show how it can set the initial conditions in a class of inflationary
theories.

\subsection{Damour Polyakov mechanism}

Scalar fields coupled to matter and with a long interaction range compared to the size of the solar system may exist and play a role in the acceleration
of the expansion of the Universe \cite{Khoury:2010xi}. This would be at odds with experimental tests on the existence of fifth forces in the solar system such as the one obtained by the Cassini probe \cite{Bertotti:2003rm}.
Fortunately, the scalar field can see its effects screened in dense environments and therefore evade gravitational tests. This is particularly the case for models subject to the
Damour-Polyakov mechanism which will be briefly reviewed in this section.

Consider a scalar tensor theory defined by the action
\be
S=\int d^4 x \sqrt{-g^E}(\frac{R_E}{16\pi G_N} -\frac{(\partial\phi)^2}{2} -V(\phi))+S_m (\psi, A^2(\phi) g_{\mu\nu}^E)
\ee
where $A(\phi)$ is an arbitrary function, $V(\phi)$ is the scalar potential and matter fields couple to the rescaled metric
\be
\tilde g_{\mu\nu}= A^2(\phi) g_{\mu\nu}^E
\ee
where the superscript E refers to the Einstein frame.
In the presence of non-relativistic matter, the dynamics of the scalar field are governed by the effective potential\footnote{We conventionally subtract the energy density $\rho_m$ from the effective potential. This is irrelevant when it comes to the dynamics of the field $\phi$ governed by the Klein-Gordon equation.}
\be
V_{\rm eff}(\phi)= V(\phi) +(A(\phi)-1) \rho_m
\ee
where $\rho_m$ is the matter energy density.
When the effective potential has a matter dependent minimum $\phi_{\rm min}(\rho_m)$, and if in dense environments the effective coupling to matter
\be
\beta(\rho_m)= m_{\rm Pl} \frac{d\ln A(\phi)}{d\phi}\vert_{\phi=\phi_{\rm min}(\rho_m)}
\ee
evaluated at the minimum, converges to zero when the density increases, gravity tests on the existence of fifth forces can be evaded. This is the Damour-Polyakov mechanism whereby
the effective coupling of the scalar field to matter becomes vanishingly small in dense environments.
The string dilaton \cite{Damour:2002mi,Damour:2002nv} at strong coupling is an example where
\be
V(\phi)=V_0 e^{-\phi/m_{\rm Pl}},\ \ A(\phi)= 1+\frac{A_2}{2m_{\rm Pl}^2} (\phi-\phi_\star)^2
\ee
and $A_2>0$ to guarantee that the field converges to $\phi_{\star}$ in dense environments where the coupling to matter vanishes.
Another family of such models are the symmetrons \cite{Hinterbichler:2010es} where $V(\phi)$ is a symmetry breaking potential where the symmetry is restored at the origin and broken for $\phi=\phi_{\rm sym}$. When the function
\be
A(\phi)= 1+\frac{A_2}{2m_{\rm Pl}^2} \phi^2
\ee
is chosen, with $A_2>0$,  the symmetry is restored in dense environments and broken in vacuum.

\subsection{Inflationary initial conditions}

We will apply the Damour-Polyakov mechanism to set the initial conditions of inflation in certain inflationary scenarios. More specifically, let us concentrate
on hilltop inflation where the inflation starts its slow roll phase close to the top of a potential situated at $\phi=\phi_\star$.
The inflaton needs to have initial conditions close to $\phi_\star$ with a very small velocity otherwise the slow roll of the inflaton towards the bottom of the hill may not be possible, e.g.  the inflaton being
in fast roll. This does not happen if there exists an initially dominant energy density of non-relativistic matter
\be
\rho_m= \frac{\rho_0}{a^3}
\ee
where the scale factor is chosen to be $a_0=1$ initially.  In this case,
the effective potential close to $\phi_\star$ reads
\be
V_{\rm eff}(\phi)= V(\phi_\star) + (\frac{A_2\rho_m}{m_{\rm Pl}^2}+ m_\star^2)\frac{(\phi-\phi_\star)^2}{2}+\dots
\ee
with a minimum at $\phi=\phi_\star$ which is stable when the effective mass of the scalar field at $\phi_\star$ is positive.
This is the fulfilled  when the condition
\be
\frac{A_2\rho_m}{m_{\rm Pl}^2} \ge \vert m_\star ^2\vert
\label{mass}
\ee
is met.
The negative mass of the bare potential in the absence of matter is defined by  $m_\star^2=\frac{d^2 V}{d\phi^2}\vert_{\phi=\phi_\star}$. We have assumed that
 the scalar field couples to matter with a function $A(\phi)$ having a minimum for $\phi=\phi_\star$, and reading therefore
\be
A(\phi)=  1+\frac{A_2}{2m_{\rm Pl}^2} (\phi-\phi_\star)^2+\dots
\ee
in the vicinity of $\phi_\star$.
As long as  $\rho_m\gg V(\phi_\star)$,  the Hubble rate is dominated by the matter density
\be
H^2=\frac{\rho_m}{3m_{\rm Pl}^2}
\ee
and the Universe is matter dominated.
As the matter density is diluted away, the condition (\ref{mass}) is eventually violated implying that the scalar field is not stuck at the origin anymore and starts rolling down along the
potential $V(\phi)$. As this triggers the beginning of slow roll inflation\footnote{In some models, the matter density can become smaller than $V(\phi_\star)$ before the condition (\ref{mass}) is violated, implying that inflation starts before the effective mass of the inflaton becomes negative. This period is automatically very short as the matter density decreases exponentially fast in this regime. Slow roll inflation then follows. If the condition (\ref{mass}) is violated during the matter dominated era, the effective mass becomes negative but the field stays at the origin until $V(\phi_\star)$ becomes dominant and inflation starts. In both cases, when inflation starts the field acquires quantum fluctuations of order $H$ which displace the inflaton from the origin and initiate the beginning of the slow roll phase.}
the matter density is very rapidly reduced to a negligible part of the energy budget of the Universe and the effective potential coincides
with $V(\phi)$ implying that the inflationary dynamics are not impaired.

Of course, this mechanism to set the initial conditions prior to inflation requires the existence of non-relativistic matter when one may expect that radiation is more likely. In fact, if the models of inflation are embedded in high energy models, we expect that many non-relativistic particles may be present. In grand unified models, this is the case of the very massive Higgs fields used in the breaking of the grand unified gauge group. Similarly, in string theory one expects Kaluza-Klein towers of extremely massive states. It is not unlikely that at lower energy some of these particles may be present and serve as the non-relativistic background of matter prior to inflation. This would certainly be enough to set the initial conditions of inflation as we have presented it.
This early Damour-Polyakov mechanism works for any hilltop model as we have seen. In particular, it can be used in the symmetron situation where a gauge symmetry is restored prior to inflation. We will apply this fact to conformal inflation later in this paper in the context of Higgs inflation.

\section{Conformal Inflation}

\subsection{Effective field theory}

In this section we first set up an effective field theory for conformal inflation.  Firstly we consider the
theory without ordinary matter, which will be discussed in a later section.
We consider a multi-field model with fields $\phi^i$, $i=1\dots N+1$, which is invariant under the global dilation symmetry
\be
\phi^i \to \lambda \phi^i, \ \ g_{\mu\nu} \to \frac{g_{\mu\nu}}{\lambda^2}
\ee
and whose action is
\be
S_\phi= \int d^4 x \sqrt{-g} ( f(\phi^i) R - h_{ij} \partial_\mu \phi^i \partial^\mu \phi^j -\tilde V(\phi^i)).
\ee
where $h_{ij}$ is a constant matrix.
The coupling function $f(\phi^i)$ is of degree two and the bare potential $\tilde V(\phi^i)$ is of degree four under dilations determined by
$\lambda$,  a real number which can be promoted to become a space-time dependent field. For that,
it is convenient to factor out the dilaton field $\rho$ by defining
\be
\phi^i= \rho z^i, \ \ g_{\mu\nu}= \frac{1}{\rho^2} \tilde g_{\mu\nu}.
\ee
In the following we shall be using the Ricci scalar in a different frame and their relation:
\be
R= \rho^2 (\tilde R +6\tilde g^{\mu\nu} \tilde D_\mu \tilde D_\nu \ln \rho -6 \tilde g^{\mu\nu} \partial_\mu \ln\rho \partial_\nu \ln\rho)
\ee
where $\tilde D_\mu$ is the covariant derivative associated with $\tilde g_{\mu\nu}$,
which allows one to rewrite
the Lagrangian as
\be
S_\phi=\int d^4 x \sqrt{-\tilde g} ( f(z^i) \tilde R -6 \partial^\mu f(z^i) \partial_\mu \ln \rho - 6f(z^i) (\partial\ln\rho)^2 - \frac{h_{ij}}{\rho^2} \partial_\mu(\rho  z^i) \partial^\mu (\rho z^j) - \tilde V(z^i))
\ee
where we have used one integration by parts.
This action is a scalar-tensor theory where the field dependent Newton constant $1/f(z^i)$ can be made truly constant by another change in field space.
This redefinition allows one to go to the Einstein frame where the metric becomes
\be
\tilde g_{\mu\nu}=\frac{1}{f(z^i)}  g_{\mu\nu}^E
\ee
implying that the action is redefined as:
\be
S_{\rm eff}=\int d^4 x \sqrt{-g^E} ( R^E - 6(\partial \ln \rho)^2 -6 \partial^\mu  \ln f\partial_\mu \ln \rho -\frac{3}{2}(\partial \ln f)^2-\frac{\tilde h_{ij}}{\rho^2} \partial_\mu(\rho  z^i) \partial^\mu (\rho z^j) -  V(z^i)).
\ee
We have  introduced the new metric in field space and the potential:
\be
\tilde h_{ij}= \frac{h_{ij}}{f(z^i)}, \  V(z^i)=\frac{\tilde V(z^i)}{f^2(z^i)}
\ee
Notice that in this frame the Planck scale is conventionally set to a pure number equal to $\sqrt 2$.

One crucial feature of the action in the Einstein frame, which follows from scale invariance, is that the
 $\ln\rho$ field, i.e. the dilaton,  appears only through its derivatives. This immediately leads to the existence of a conservation equation which can be derived from the Klein-Gordon equation for $\ln\rho$:
\be
D_\mu J^\mu=0
\ee
where the conserved current is found to be:
\be
J^\mu = \frac{1}{\rho^2 f(z^i)}\partial^{\mu} (\rho^2 ( h(z^i) + 6 f(z^i)))
\ee
We have introduced the quadratic form:
$
h(z^i)= h_{ij} z^i z^j.
$
So far in defining the dilaton field $\rho$ we have not specified how the rescaled fields $z^i$ have been normalised.
Here we choose to concentrate on field configurations such that:
\be
h(z^i) + 6 f(z^i)=1.
\label{z}
\ee
In this case and requiring a vanishing value for the current $J^\mu=0$,  implies that
\be
\rho=\rho_0
\ee
and the dilaton is a constant.
For these configurations the effective action becomes
\be
S_{\rm eff}=\int d^4 x \sqrt{-g} ( R  -\frac{3}{2}(\partial \ln f)^2-\tilde h_{ij} \partial_\mu  z^i \partial^\mu  z^j -  V(z^i))
\ee
Using the constraint (\ref{z}), we have simplified the action and reduced it to a scalar model involving $(N+1)$ constrained fields. 

\subsection{Inflationary Dynamics}

We will simplify the general models that we have obtained by assuming that the matrices $h_{ij}$ and $f_{ij}$ commute. In this case, they can be both diagonalised in the same basis with eigenvalues
$f_i$ and $h_i$.
The constraint on the scalar field becomes in this preferred basis
\be
\sum_{i=1}^{N+1} (6f_i + h_i) z_i^2=1
\ee
which parameterises a unit sphere in terms of the rescaled field\footnote{We focus on models where both $f_i$ and $h_i$ are positive. Extensions where some of these entries
are negative leading to hyperbolic spheres are left for future work.}
\be
u^i(\theta^\alpha)= \frac{z^i}{\sqrt{6f_i+h_i}}
\ee
where
$\sum_{i=1}^{N+1}(u^i(\theta^\alpha))^2=1$.
The fields $\theta_\alpha, \alpha=1\dots N,$  parameterising this sphere define the low energy dynamics of the model which are nothing but a $\sigma$-model with the kinetic terms
\be
{\cal L}_{kin}=g_{\alpha\beta} \partial_\mu \theta^\alpha\partial^\mu \theta^\beta
\ee
where the metric is given by
\be
g_{\alpha\beta}= (\frac{3}{2}\frac{\partial_{z^i} f \partial_{z^j} f}{f^2} + \frac{h_{i}\delta_{ij}}{f})\frac{1}{\sqrt{6f_i+h_i}}\frac{1}{\sqrt{6f_j+h_j}}\frac{\partial u^i}{\partial \theta^\alpha} \frac{\partial u^j}{\partial \theta^\beta}
\ee
at low energy.

Let us now consider sufficient conditions for inflation in these models.
First of all, we consider a field configuration on the sphere defined by $\theta_0^\alpha$ which maximises both the coupling function $f(\theta^\alpha)$ and the bare potential $\tilde V(\theta^\alpha)$.
In the vicinity of this point, the metric defining the kinetic terms reduces to
\be
g_{\alpha\beta}^{(0)}\approx \frac{h_{i}\delta_{ij}}{f^{(0)}}\frac{1}{{6f_i+h_i}}\frac{\partial u^i}{\partial \theta^\alpha}\vert_0 \frac{\partial u^j}{\partial \theta^\beta}\vert_0
\ee
where the expression is evaluated at $\theta^\alpha_0$.
Expanding $\tilde V$ and $f$ as
\be
\tilde V (\theta^\alpha) \approx V_0(1 -  v_{\alpha\beta}(\theta^\alpha -\theta_0^\alpha)(\theta^\beta -\theta_0^\beta))
\ee
and
\be
\tilde f (\theta^\alpha) \approx f^{(0)}(1 - \frac{1}{2} f_{\alpha\beta}(\theta^\alpha -\theta_0^\alpha)(\theta^\beta -\theta_0^\beta))
\ee
where both $v_{\alpha\beta}$ and $f_{\alpha\beta}$ are positive definite matrix,
the potential in the vicinity of the maximum of $f$ and $\tilde V$ becomes now
\be
V (\theta^\alpha)\approx \frac{V_0}{f^{(0)}}( 1+ (f_{\alpha\beta} - v_{\alpha\beta})(\theta^\alpha -\theta_0^\alpha)(\theta^\beta -\theta_0^\beta)).
\ee
The rescaled mass matrix
\be
m^2_{\alpha\beta}= f_{\alpha\beta} - v_{\alpha\beta}
\ee
is not positive definite anymore. Let us diagonalise this matrix and assume at first that there are no zero eigenmasses. We will deal with these massless states in the next section when they are associated with Goldstone directions.  In the diagonal basis the eigenmasses $m_\alpha^2$ which are strictly positive are associated with attractive directions in field space.
Along these directions, the fields settle at the maximum of $f$ and $\tilde V$ under the cosmological evolution. We focus on the directions such that $m^2_\alpha <0$.
The fields starting very close to the maximum of $f$ and $\tilde V$ will eventually roll down along these eigendirections  where the effective action reduces to
\be
{\cal L}_{\rm inf}= \bar g_{\alpha\beta} \partial_\mu V^\alpha \partial^\mu V^\beta +\frac{V_0}{f^{(0)}}( 1+ m^2_\alpha (V^\alpha)^2)
\ee
where the basis of relevant fields is given by
\be
V^\alpha= O^{\alpha \beta}(\theta^\beta -\theta_0^\beta),\ \ \bar g_{\alpha\beta}= O^{\gamma\alpha} g_{\gamma\delta }^{(0)}O^{\delta\beta}
\ee
and the mass matrix is diagonalised by the orthogonal transformation $O^{\alpha\beta}$.
This defines a multifield inflationary model of the hilltop type where the inflaton starts close to the maximum of the potential and  rolls down along the
directions with negative mass squared.

The dynamics are described by the Friedmann equation
\be
H^2= \frac{1}{6} ( \bar g_{\alpha\beta} \dot V^\alpha \dot V^\beta + \frac{V_0}{f^{(0)}}( 1+ m^2_\alpha (V^\alpha)^2)
\ee
and the Klein-Gordon equation
\be
\bar g_{\alpha\beta} ( \ddot V^\beta + 3H \dot V^\beta)= - \frac{V_0}{f^{(0)}} m_\alpha^2 \delta_{\alpha\beta} V^\beta
\ee
from which one deduces the evolution of the Hubble rate
\be
\dot H=-\frac{1}{2} \bar g_{\alpha\beta} \dot V^\alpha \dot V^\beta.
\ee
Slow roll inflation occurs close to $V^\alpha=0$ where the Klein-Gordon equation
reduces to
\be
3 H \bar g_{\alpha\beta}\dot V^\beta=-\frac{V_0}{f^{(0)}} m_\alpha^2 \delta_{\alpha\beta}V^\beta
\ee
where
\be
H^2=\frac{V_0}{6f^{(0)}}
\ee
 to leading order.
This is true as long as $\vert \frac{\dot H}{H^2}\vert \ll 1$ which is guaranteed when  the kinetic energy is negligible compared to $H^2$, i.e. close enough to the origin.
Similarly the slow roll evolution of $V^\alpha$ stands as long as
\be
\bar g_{\alpha\beta}  \ddot V^\beta= -(\frac{m^2_\alpha}{3H}\delta_{\alpha\beta}  +\frac{\dot H}{H} \bar g_{\alpha\beta}) \dot V^\beta,
\ee
as obtained from the slow roll equation, is smaller than $3H \bar g_{\alpha\beta}\dot V^\beta$. As the second term on the right hand side is small due to the smallness of the kinetic energy,
this condition is met provided that
\be
\vert \bar g_{\alpha\beta}\dot V^\beta\vert \gg \vert \frac{m^2_\alpha}{H^2} \delta_{\alpha\beta} \dot V^\beta\vert
\ee
along the inflationary trajectories. Taking $\dot V^\alpha$ along the eigenvectors of $\bar g_{\alpha\beta}$ where the eigenvalues are $\bar g_\alpha$, this condition is automatically realised provided
\be
\max (\vert \frac{m^2_\alpha}{H^2}\vert ) \ll \min (\bar g_\alpha)
\ee
which generalises the condition on the second slow roll parameter $\eta= 2\vert \frac{V''}{V}\vert \ll 1 $ for canonically normalised single field inflationary models.

Inflation is maintained as long as these conditions are fulfilled. Generically as the field rolls away from the top of the potential, the kinetic energy increases and the curvature too implying that slow roll eventually breaks down. Determining when this happens requires more than the simple approximations we have used here and will be analysed in full detail when we revisit the Higgs-Dilaton models.

\subsection{Gauged models}

We are also interested in inflationary models where the scalar fields are charged under a compact gauge group $G$. We take the $(N+1)$ scalar fields to belong to a representation ${\cal R}$ of the gauge group. This representation is taken to be a reducible one in the general case. As we have focused on real scalar fields, we assume that the representation ${\cal R}$ is real although the complex extension of our results can be easily envisaged. Gauge invariance requires that the coupling function $f(\phi^i)$ and the potential $\tilde V(\phi^i)$ are invariant under $G$. As the real symmetric matrices $f_{ij}$ and $h_{ij}$ commute, they can be diagonalised simultaneously.
The diagonalisation implies that one can decompose the space of fields
\be
{\mathbb{R}}^{N+1} = \oplus_{i=1}^p V_i
\ee
where each subvector space $V_i$ corresponds to the eigenspaces of  $h_{ij}$ and $f_{ij}$. Moreover, Schur's lemma guarantees that $V_i$ is left invariant by the gauge group $G$. This implies that the quadratic forms $f(z^i) = f_{ij} z^i z^j$ and $h(z^i)= h_{ij} z^i z^j$ are left invariant by the action of the gauge group $G$. Hence the constraint on the fields $6 f(z^i) + h(z^i)=1$ is invariant under $G$.
Gauge invariance is preserved in the effective action depending on the fields $z^i$.

Inflation occurs in the vicinity of a maximum of both $f$ and $\tilde V$ where the vacuum expectation value of $\phi^i$ breaks the gauge symmetry down to a subgroup $H\subset G$.
After symmetry breaking, we have $\dim G/H$ Goldstone modes which correspond to mass eigenvalues of the mass matrix $m_\alpha=0$. By the Higgs mechanism, these Goldstone fields become the longitudinal polarisation
of the massive gauge bosons associated with the breaking of the gauge group. In this case, we are left with massive fields which can lead to inflation in the case of negative and small enough squared masses.

As a useful example, let us consider the gauge group $SO(N)$ acting on the $(N+1)$ fields as the representation $N\oplus 1$. The singlet is associated with the dilaton. Assume that gauge symmetry is broken down to $SO(N-1)$ by the vacuum during inflation.
Then we have $(N-1)$ Goldstone directions which are eaten by the Higgs mechanism and one remaining field. This field is a natural candidate for inflation. This scenario is explicitly realised in the Higgs-Dilaton models.

\subsection{Damour-Polyakov mechanism}

When scale invariance is restricted to the inflationary sector and is explicitly broken by a matter sector coupled to gravity in a minimal way, the effective potential in the Einstein frame is modified. This is due to the presence of non-relativistic matter whose mass breaks conformal invariance and lifts the dilatonic flat direction. This changes the dynamics of all the fields. Here we take the total action to be
the inflationary one together with  matter, as given by the action:
\be
S= S_\phi + S_m (\psi^a, g_{\mu\nu})
\ee
where $S_m$ is the matter action for some matter fields $\psi^a$.
In the Einstein frame, after redefining the metric as before,
\be
g_{\mu\nu}= f^{-1}(\phi^i) g^E_{\mu\nu}
\ee
the matter action becomes
\be
S_m (\psi^a, \frac{g^E_{\mu\nu}}{f(\phi^i)}).
\ee
The dilaton field $\rho$  acquires a potential due to the presence of non-relativistic matter. This lifts the flat direction that we have previously obtained as the dilatonic current $J^\mu$  is not conserved anymore. Using homogeneity
$f(\phi^i)=\rho^2 f(z^i)$, the matter density $\rho_m$ induces the potential term $\rho^{-1} f^{-1/2}(z^i) \rho_m$ leading to the effective scalar potential
\be
V_{\rm eff}(z^i)=  V(z^i) + \frac{\rho_m}{\rho \sqrt{f(z^i)}}
\ee
Notice that the dilaton now acquires a potential and therefore evolves with time.
Assuming that $f(z^i)$ has a unique maximum, we see that the matter dependent term has a minimum there. When the matter density is high, this term is enough to attract the fields $z^i$ to the maximum of $f$. More precisely, expanding the effective potential, we get
\be
V_{\rm eff}(\theta_\alpha,\rho)=  \frac{V_0}{f^{(0)}}( 1+ m^2_{\alpha\beta}(\theta^\alpha -\theta_0^\alpha)(\theta^\beta -\theta_0^\beta))+ \frac{\rho_m}{\rho(f^{(0)})^{1/2}}(1+\frac{1}{4} f_{\alpha\beta}(\theta^\alpha -\theta_0^\alpha)(\theta^\beta -\theta_0^\beta))
\ee
The  effective mass of the fields is now
\be
M^2_{\alpha\beta}= (m^2_{\alpha\beta} + \frac{(f^{(0)})^{1/2}} {4V_0} \frac{\rho_m}{\rho}f_{\alpha\beta})\frac{V_0}{f^{(0)}}
\ee
As long as $\rho_m/\rho$ is large enough to offset the negative eigenvalues of $m^2_{\alpha\beta}$, the mass matrix becomes positive definite and proportional to $f_{\alpha\beta}$.
When the matter density dominates, i.e. $V_0\lesssim {(f^{(0)})^{1/2}} \frac{\rho_m}{\rho}$, the Friedmann equation reads
\be
H^2\approx \frac{\rho_m}{6\rho(f^{(0)})^{1/2}}
\ee
and the mass matrix of the matter fields is
\be
M^2_{\alpha\beta}\approx \frac{3}{2} H^2 f_{\alpha\beta}.
\ee
The fields $\theta^\alpha$ oscillate around $\theta^\alpha_0$ and converge to this attractor like $a^{-3/2}$. This sets the initial conditions for inflation. As
soon as the matter density term is redshifted enough and does not dominate anymore, the fields start evolving and slow roll inflation starts. When this happens, the matter density becomes very rapidly negligible and
the conformal scenario for inflation takes place.

\section{Late Time Dynamics}

\subsection{Coupling to the standard model}

When inflation stops, the fields roll fast towards the absolute minimum of the potential $\tilde V(\phi^i)$ which must have zero value due to scale invariance. At this point, the fields $z^i$ have a vacuum expectation value which gives a mass to particles of matter coupled to the conformal inflationary sector. We have already seen that non-relativistic matter which may have been present prior to inflation has been washed away implying that only matter acquiring mass at the end of inflation during the reheating process has any chance of being still around in the late time Universe. We focus on conformal inflation models where the scalar fields are charged under a
gauge group $G$ and consider such a matter sector which may be thought of as a simplified version of the standard model. Let us generically denote by $A_\mu$ the gauge fields and $\psi^a$ the matter fermions.
We assume that the matter Lagrangian comprises the Yukawa couplings
\be
S_{\rm SM}= \int d^4x \sqrt{-g}( i\bar\psi^a \gamma^\mu D_\mu \psi^a + \lambda_{iab} \phi^i \bar \psi^a\psi^b)
\ee
where $D_\mu$ includes the gauge and spin connections. The Yukawa couplings $\lambda_{iab}$ are gauge invariants under the action of $G$.
The gauge fields have the usual gauge kinetic terms
\be
S_{\rm gauge}= -\frac{1}{4} \int  d^4x \sqrt{-g} F^2
\ee
where $F_{\mu\nu}$ is the gauge field strength.
Let us consider the action of the rescaling by the dilaton where the matter fields transform as
\be
\psi^a= \rho^{3/2} \tilde \psi^a
\ee
and the gauge fields are invariant $\tilde A_\mu =A_\mu$.
The matter action becomes
\be
S_{\rm SM}= \int d^4x \sqrt{-\tilde g}(i \bar{\tilde\psi}^i \tilde \gamma^\mu \tilde D_\mu \tilde \psi^i + \lambda_{ijk} z^i \bar{\tilde \psi}^j\tilde \psi^k)
\ee
where the spin connection and the gamma matrices have been transformed according to the change of metric.
Similarly, the gauge action is invariant under this field rescaling.
As a result, the dilaton field $\rho$ is absent from the coupling of the $z^i$ fields to matter.

Let us now go to the Einstein frame while redefining the fermions fields
\be
\tilde \psi^a= f^{3/4}(z^i)\psi^a_E
\ee
implying that the matter action becomes
\be
S_{\rm SM}= \int d^4x \sqrt{- g_E}( i\bar\psi_E^a \gamma_E^\mu  D^E_\mu  \psi_E^a + \lambda_{iab} \frac{z^i}{f^{1/2}(z^i)}  \bar \psi_E^a \psi_E^b).
\ee
In this frame, the Planck mass is equal to $\sqrt 2$. In physical units, the fermionic mass matrix is
\be
m_{ab}^\psi= \lambda_{iab} \frac{<z^k>}{\sqrt 2f^{1/2}(<z^i>)} m_{\rm Pl}
\ee
once the fields $z^i$ have acquired a vev.
This gives a mass to the fundamental particles akin to the quarks and leptons.
Bound states such as the proton and the neutron acquire their mass after the QCD transition where the gluons condense. Here in the Einstein frame, the gauge kinetic terms do not depend on either $\rho$ or $z^i$
implying that bounds states have  scalar-independent masses.

The gauge bosons corresponding to the broken part of the gauge group $G$ acquire a mass coming from the gauged kinetic terms
\be
\int d^4x \sqrt{-g} h_{ij} D_\mu \phi^iD^\mu \phi^j
\ee
 which leads to the mass term for the gauge bosons $A_\mu$
\be
\int d^4x\sqrt{-g} g_G^2\rho^2 \bar h (<z^i>) A^\mu A_\mu
\ee
where $g_G$ is the gauge coupling and $\bar h_{ij}$ is the restriction of $h_{ij}$ to the subspace of the charged scalar fields breaking the gauge group $G\to H$, and $\bar h (<z^i>)=\bar h_{ij} <z^i> <z^j>$.
The Einstein frame mass term reads
\be
\int d^4x \sqrt{-g_E} g_G^2 \frac{\bar h (<z^i>)}{f (<z^i>)} A^\mu A_\mu
\ee
corresponding to a physical mass for the gauge fields
\be
M^2_A=  g_G^2 \frac{\bar h (<z^i>)}{ f (<z^i>)}m_{\rm Pl}^2
\ee
which requires a hierarchy $\bar h (<z^i>) \ll f (<z^i>)$ to ensure that gauge fields have a mass much lower than the Planck scale.

\subsection{Conformal models in unimodular gravity}

We have seen that the dilaton field decouples from the dynamics of the conformal models as soon as non-relativistic matter particles see their energy density washed away during inflation.
At low energy, scale invariance is broken by the vev's of the fields $<z^i>$ and both fermions and gauge bosons acquire masses which do not depend on the dilaton field at tree level.
It is tempting to use the vanishing of the potential energy $V(<z^i>)=0$ at the minimum of the potential as a starting point to explain the late time acceleration of the Universe. Unfortunately, the vanishing of vacuum energy is spoilt at low energy whenever matter particles become non-relativistic and are integrated out of the dynamics. The vacuum energy receives large contributions which are quartic in the particle masses and independent of the dilaton field. To alleviate this problem, we will embed the conformal models into a unimodular gravity setting. As recalled in the introduction, this has the advantage of guaranteeing that particle physics contributions to the vacuum energy do not gravitate and do not influence the dynamics of the model. On the other hand, unimodular gravity breaks scale invariance explicitly in a very mild manner, introducing one energy scale which plays the role of a cosmological constant in the Jordan frame. This constant is set by the initial conditions of the model in the early Universe and not by the decoupling of matter particles. In all the models that we have discussed and in the Einstein frame, the breaking of scale invariance in unimodular gravity implies that the dilaton acquires  a potential which can drive dark energy.

More precisely, unimodular gravity is a modification of General Relativity where the cosmological constant is not dynamical anymore, therefore solving the cosmological constant problem. This follows from the fact that the cosmological constant does not couple to the metric tensor
\begin{equation}
S= \int {\rm d}^4x (\frac{R}{2\kappa_4^2} + \Lambda_0^4)
\label{Uni}
\end{equation}
where $\det g=-1$. The cosmological constant $\Lambda_0^4$ is non-dynamical. In Einsteinian gravity, the cosmological constant receives quantum field corrections due to the non-trivial nature of matter vacuum fluctuations. In unimodular gravity,  the  radiative corrections do not lead to the acceleration of the universe despite the possible existence of a large cosmological constant.  Of course, this possibility does not explain the observed acceleration of the late time universe. All it provides is a mechanism for the irrelevance of the quantum fluctuations of the vacuum. Therefore another mechanism needs to be unveiled  to understand the late time acceleration of the universe.

Let us embed the conformal models that we have introduced in the framework of unimodular gravity using the action
\begin{equation}
S_{\rm uni}= \int {\rm d}^4x (f(\phi^i) {R} -h_{ij} \partial_\mu{\phi^i}\partial^\mu{\phi^j} -  \tilde V(\phi^i)).
\end{equation}
In this frame and coupling matter minimally, radiative corrections due to the presence of matter only, i.e. involving no gravitons, lead to a non-gravitating cosmological constant with no relevance
on the dynamics of the theory. Moreover,
it is simpler to describe unimodularity   using a Lagrange multiplier
\begin{equation}
S= \int {\rm d}^4x [\sqrt{-g}(f(\phi^i) {R} -h_{ij} \partial_\mu{\phi^i}\partial^\mu{\phi^j} -  \tilde V(\phi^i)) -\Lambda^4 (\sqrt {-g}-1))]
\end{equation}
where $\Lambda$ can be shown to be constant \cite{Shaposhnikov:2008xb}. The first part of the action is scale invariant while the one involving $\Lambda$ is not. It turns out that
the equations of motion   coincide with
the  Einstein  and the Klein-Gordon equations of  the fully covariant Lagrangian
\begin{equation}
{\cal L}= \sqrt{-g}[f(\phi^i) {R} -h_{ij} \partial_\mu{\phi^i}\partial^\mu{\phi^j} -  \tilde V(\phi^i) -\Lambda^4].
\end{equation}
The dynamics of the theory are determined by this scalar-tensor theory which is scale invariant  if and only if $\Lambda=0$.

In these models, one can introduce the dilaton $\rho$ which decouples from the scalar part of the action and appears explicitly in the $\Lambda^4$ term. If $\Lambda^4$ is small enough, it plays no role at high energy during inflation
and only becomes relevant at low energy where it may drive the acceleration of the Universe.
Let us write the effective action at low energy  when the fields $z^i$ have settled at the minimum of the potential $V(z_i)$ whose value vanishes there, i.e we put $\phi^i= \rho <z^i>$ and use $\tilde V(<z^i>)=0$.
We find that the effective action for the dilaton is
\begin{equation}
{S_{\rm eff}}= \int d^4x \sqrt{-g}[(f_0\rho^2  {R}- h_0 (\partial \rho)^2 - \Lambda^4]
\end{equation}
where we have identified $  f(<z^i>)=f_0$ and $h(<z^i>)=h_0$.
In  order to analyse the gravitational and cosmological properties
of the model, it is easier to go to the Einstein frame with
$
g_{\mu\nu}= A^2 (\rho) g_{\mu\nu}^E
$
where
$
A^{-2}(\rho)=f_0  \rho^2.
$
It is useful to normalise the dilaton field as
\be
\rho=\frac{1}{\sqrt {f_0}} e^{\sqrt f_0 \phi/\sqrt{2}}
\ee
leading to the effective action
\begin{equation}
 S= \int {\rm d}^4 x \sqrt{-g^E}[  {R^E} - \frac{1}{2} \partial_\mu \phi\partial^\mu \phi -\Lambda^4 e^{-4\frac{\sqrt{f_0}}{\sqrt{2}}\phi}]
\end{equation}
This action depends only on the parameter   $f_0$. The model reduces to a quintessence model with an exponential potential. The dilaton field is not a flat direction anymore but has a potential whose origin results from
the breaking of scale invariance in unimodular gravity.
The only parameter which contains some information on the early dynamics of the model is $f_0$ which depends on the functional form of $f(\phi^i)$.

\subsection{Cosmological and gravitational properties}

At late time, when the matter density has decreased enough, the potential term for the dilaton starts dominating and the dark energy scale is given by the value of the dilaton now
\be
\Lambda^4 e^{-4\frac{\sqrt{f_0}}{\sqrt{2}}\phi_0}\approx 6\Omega_{\Lambda 0} H_0^2
\ee
where we have used reduced units where the Planck scale is equal to $\sqrt 2$, $H_0$ is the Hubble rate now and $\Omega_{\Lambda 0}\approx 0.73$ is the dark energy fraction.
When dark energy dominates, the model is known to have an attractor solution for the dilaton and that its equation of state, the ratio between the pressure and the energy density is given by
\begin{equation}
1+w= -\frac{16f_0}{3}
\end{equation}
where $p_\phi= \dot\phi^2/2 -V(\phi)$, $\rho_\phi=\dot\phi^2/2 +V(\phi)$ and $w=p_\phi/\rho_\phi$.
Of course, dark energy is not completely dominant now as $\Omega_{\Lambda 0}\ne 1$, but this is a reasonable order of magnitude. This forces $f_0$ to be no more than $0.1$ to comply with recent data.
We can calculate the mass of the dilaton field now
\be
m^2_\rho= 8f_0\Lambda^4 e^{-4\frac{\sqrt{f_0}}{\sqrt{2}}\phi_0}
\ee
identified with
\be
m^2_\rho\approx 48f_0 \Omega_{\Lambda_0} H_0^2
\ee
which is of the order of the Hubble rate and would therefore imply that the dilaton mediates a very long range scalar interaction, of range similar to  the size of the observable Universe.
We have shown that the leptons and the bound states of QCD have masses which are dilaton-independent
\be
\frac{d\ln m_\psi}{d\rho}=0
\ee
thanks to conformal invariance at the classical level. For such fermions, and when they have  acquired masses as the fields $z^i$ are stabilised at the vacuum expectation value $<z^i>$, the effective Lagrangian becomes
\begin{equation}
{S_{\rm eff}}= \int d^4x \sqrt{-g}[(f_0\rho^2  {R}- h_0 (\partial \rho)^2 -i\bar\psi \gamma^\mu D_\mu \psi-m_\psi\rho \bar\psi \psi]
\end{equation}
where $h_0=1-6f_0$ is close to one. Denoting by $h_{\mu\nu}$ the normalised graviton such that $g_{\mu\nu}= \eta_{\mu\nu}+ \frac{h_{\mu\nu}}{\sqrt{f_0} <\rho>}$, the coupling of the perturbed field $\delta \rho$ to the Ricci scalar is symbolically of the form $  \frac{\delta \rho}{<\rho>} \tr (\partial h)^2$ where $<\rho>$ is the background value around which the perturbative expansion is performed. Similarly, the fermionic mass term induces a coupling between the graviton and the fermions of the form $\frac{m_\psi}{f_0<\rho>}  \tr (h^2) \bar\psi\psi $. At one loop, the diagrams
with one external $\delta \rho$ and two fermions $\bar\psi\psi$ can be drawn by joining the vertices with two graviton lines using the  propagators $1/p^2$. The resulting Feynman diagram is quadratically divergent and can be estimated by putting a cut-off scale at the symmetry breaking scale where fermion acquire a mass $g_G \sqrt{\bar h} <\rho>$. The result is of order $g_G^2 \frac{\bar h}{f_0} m_\psi \delta\rho \bar\psi \psi$. Going to the Einstein frame by redefining the Planck scale yields an effective coupling of the form
\be
g_G^2 \frac{\bar h}{f_0} m_\psi \frac{\delta\rho}{\sqrt{f_0}<\rho>} \bar\psi_E \psi_E \approx g_G^2 \frac{\bar h}{\sqrt 2 f_0} m_\psi \delta\phi \bar\psi_E \psi_E
\ee
corresponding to a coupling between the dilaton $\phi$ and fermions of order
\be
\beta_\phi \approx g_G^2 \frac{\bar h}{\sqrt 2 f_0}= \frac{\sqrt 2}{2}  \frac{M_A^2}{m_{\rm Pl}^2}
\ee
which is nothing but the ratio of the gauge field masses to the Planck scale. As a result, we find that the dilaton remains almost decoupled from matter as long
as particle masses are much lower than the Planck scale.

In conclusion, we have found that the dilaton can drive the acceleration of the Universe at late time in unimodular gravity. Moreover its coupling to matter is heavily suppressed guaranteeing that
gravity tests in the solar systems are evaded.

\section{A case Study: Higgs-Dilaton Inflation}

\subsection{Higgs-Dilaton inflation}

In this section we specialise to the case of Higgs-dilaton inflation \cite{GarciaBellido:2011de}, building on our formalism of the previous section.
Let us now consider a model with $(N+1)$ fields and an invariance under an $SO(N)$ group. The symmetry is broken down to $SO(N-1)$ by the initial conditions of inflation  leading to the existence of $(N-1)$ Goldstone bosons.
We gauge this symmetry and the Goldstone modes are eaten by the Higgs mechanism where they become the longitudinal polarisations of the gauge fields.  In the standard model, the gauge group is $SU(2) \approx SO(3)$ corresponding to $N=3$. The remaining dynamical field after the Higgs mechanism is a candidate for the inflaton field for particular choices of the functions $f$ and $\tilde V$.

In the Higgs-Dilaton models, the potential is chosen to be:
\be
\tilde V(\phi^i)= \lambda (\sum_{i=1}^N (\phi^i)^2 -\alpha (\phi^{N+1})^2)^2
\ee
and the coupling to the Ricci scalar is:
\be
f(\phi^i)= \xi_h \sum_{i=1}^N (\phi^i)^2 +\xi_\phi (\phi^{N+1})^2
\ee
with the diagonal metric
\be
h_{ij}=\delta_{ij}
\ee
for the kinetic terms.

The fields are normalised such that
\be
\sum_{i=1}^{N+1} (1+6\xi_i) (z^i)^2=1
\ee
where we define $\xi_i=\xi_h$, $i=1\dots N$, and $\xi_{N+1}= \xi_\phi$.
It is convenient to introduce the parameterisation for the fields
\be
z^i= \cos \theta \frac{e^i}{\sqrt{1+6\xi_h}}
\ee
where the $N$ fields $e^i$ satisfy the constraint
\be
\sum_{i=1}^N (e^i)^2=1
\ee
 defining $(N-1)$ independent fields.
Finally we also choose:
\be
z^{N+1}= \frac{\sin\theta}{\sqrt{1+6\xi_\phi}}.
\ee
Notice that the $(N-1)$ independent $e^i$'s have no potential term, i.e. they are the massless Goldstone bosons eaten by the Higgs mechanism.
In the following we will focus on the $\theta$ field which will play the role of the inflaton.

We are now in a position to calculate the kinetic terms for the field $\theta$ using:
\be
\sum_{i=1}^{N+1} (dz^i)^2= G(\theta) d\theta^2
\ee
where we have defined
\be
G(\theta)= \frac{\sin^2\theta}{1+6\xi_h}+ \frac{\cos^2\theta}{1+6\xi_\phi}
\ee
Using this  parameterisation we find that
\be
f(\theta)=\frac{\xi_h}{1+6\xi_h} (1- (1- \frac{\tilde \xi_\phi}{\xi_h})\sin^2 \theta)
\ee
where we have introduced
$
\tilde \xi_\phi = \frac{1+6\xi_h}{1+6\xi_\phi} \xi_\phi.
$
Similarly we have that the rescaled potential $\tilde V(z^i)/f^2(z^i)$ is given by
\be
 V(\theta)= \frac{\lambda}{(1+6\xi_h)^2} \frac{(1- (1+ \tilde \alpha) \sin^2\theta)^2}{f^2(\theta)}
\ee
where
$
\tilde \alpha= \frac{1+6\xi_h}{1+6\xi_\phi} \alpha.
$
Notice that $\theta=0$ is a maximum for $f(\theta)$ and $\tilde V (\theta)$ where the gauge symmetry is broken down to $SO(N-1)$ leading to the $(N-1)$ Goldstone fields $e^i$ as expected from our general
discussion.

The potential vanishes for
\be
\sin^2\theta_0= \frac{1}{1+\tilde \alpha}
\ee
corresponding to
\be
f_0=\frac{\xi_h}{1+6\xi_h} (1- \frac{1- \frac{\tilde \xi_\phi}{\xi_h}}{1+\tilde \alpha})
\ee
This gives an order of magnitude estimate for the fermion masses
\be
m_\psi\approx \lambda_\psi \frac{\cos\theta_0}{\sqrt{\xi_h} (1- \frac{1-\frac{\tilde \xi_\phi}{\xi_h}}{1+\tilde \alpha})^1/2}
\ee
in units where the Planck scale is equal to $\sqrt{2}$  and $\lambda_\psi$ is a Yukawa coupling.
Requiring that $f_0$ is smaller than $0.1$ and $m_\psi$ is much smaller than the Planck scale can be achieved when $ \xi_h \alpha \ll \xi_\phi \ll 1$. Indeed we find
$f_0\approx \xi_\phi$ and $m^2_\psi \approx \lambda_\psi^2 \frac{\alpha}{\xi_\phi}$ which requires $\alpha \ll \xi_\phi$.

The kinetic terms for the only remaining field $\theta$ become
\be
\frac{3}{2}(\partial \ln f)^2+\tilde h_{ij} \partial z^i \partial z^j =\frac{H(\theta)}{f(\theta)} d\theta^2
\ee
which leads to the canonical term for a field $\varphi$
where have introduced
\be
d\varphi= \sqrt{2} (\frac{H(\theta)}{f(\theta)})^{1/2} d\theta.
\label{the}
\ee
The function $H(\theta)$ is such that:
\be
H(\theta)=G(\theta) + \frac{3}{2} \frac{\xi_h}{1+6\xi_h} (1-\frac{\tilde \xi_\phi}{\xi_h})^2 \frac{\sin^2(2\theta)}{1-(1-\frac{\tilde \xi_\phi}{\xi_h})\sin^2\theta}
\ee
The effective action for the field  $\varphi$ which is simply
\be
S=\int d^4 x \sqrt{-g} ( R - \frac{1}{2} (\partial \varphi)^2  - V(\varphi)).
\ee
where $ V(\varphi)=  V(\theta(\phi))$ is defined implicitly and  we have a relation $\theta(\phi)$ from (\ref{the}).
This is the inflationary model describing the effective dynamics of the Higgs-Dilaton models.

\subsection{Inflationary dynamics}
Let us analyse the dynamics of this effective model.
The field rolls slowly from the top of the potential satisfying the slow roll equation
\be
3H^2 \frac{d\varphi}{dN}= -\frac{\partial V}{\partial \varphi}
\ee
where $dN=Hdt$ is the number of e-foldings and the Friedmann equation can be approximated by
\be
3H^2= \frac{V(\varphi)}{2}
\ee
in the reduced units where the Planck scale is equal to $\sqrt 2$.
The potential as a function of $\theta$ is given by
\be
V(\theta)=\frac{\lambda}{\xi_h^2}(\frac{1-(1+\tilde \alpha) \sin^2 \theta}{1-(1-\frac{\tilde \xi_\phi}{\xi_h})\sin^2 \theta})^{2}.
\ee
We can estimate the value of the field at the end of inflation $\varphi_{\rm end}$ by requiring that the slow roll parameter $\vert \epsilon_{\rm end}\vert =1$. The first slow parameter is here
\be
\sqrt{\epsilon} \equiv \vert\frac{V_\varphi}{V}\vert
\ee
which turns out to be
\be
\sqrt \epsilon= \frac{\sqrt 2 (\tilde \xi_\phi +\xi_h \tilde \alpha)}{\sqrt{1+6\xi_h}}\frac{\vert \sin 2 \theta\vert}{\sqrt{H(\theta)} (1-(1+\tilde\alpha)\sin^2\theta)(\xi_h\cos^2\theta +\tilde \xi_\phi \sin^2\theta)^{1/2}}.
\ee
Inflaton ends close to the minimum of the potential where $(1+\tilde \alpha )\sin^2 \theta_0=1$ for a value where
\be
\cos^2 \theta_{\rm end} = 4\sqrt{3} {\xi_\phi}.
\ee
 We have assumed here that $\xi_h\gg 1,\ \xi_\phi\ll 1,\ \alpha\ll 1$ and $\xi_\phi\gg \xi_h \alpha$ as  required by the COBE normalisation, late time physics and the electroweak transition (see later for the COBE constraint).

The slow roll evolution is governed by the differential equation:
\be
\frac{d\theta}{dN}= \frac{2(\xi_\phi+\alpha \xi_h)}{1+6\xi_\phi}\frac{\sin 2\theta}{H(\theta) (1-(1+\tilde\alpha)\sin^2\theta)}.
\ee
During inflation, $H(\theta)\sim 1$ and this can be simplified:
\be
\frac{d\theta}{dN}\sim 4\xi_\phi\frac{\sin \theta}{\cos\theta}
\ee
implying that the time evolution of the inflaton is given by:
\be
\sin\theta= \sin\theta_{\rm end} e^{-4\xi_\phi  \Delta N}
\ee
where $\Delta N$ is the number of e-foldings from the end of inflation.
In this regime we have
\be
\epsilon_N\approx \frac{12 \xi_\phi^2}{\sinh^2 (4\xi_\phi \Delta N)}.
\ee
The second slow roll parameter $\epsilon_2= \frac{d\ln \epsilon_N}{2d\Delta N}$ is given by
\be
\epsilon_2= -4\xi_\phi \coth (4\xi_\phi \Delta N).
\ee
The spectral index  $n_s-1= 2\epsilon_2 - 2 \epsilon$ reduces to
\be
n_s-1= -\frac{2}{\Delta N}
\ee
in the cases when $\xi_\phi \Delta N \ll 1$ and the $1/(\Delta N)^2$ term has been dropped. For $\Delta N=60$ this is $n_s-1\approx 0.97$. Similarly the tensor to scalar ratio is given by $r= 12/(\Delta N)^2$
which is smaller than the present day observational sensitivity.

The amplitude of the CMB fluctuations fixes the COBE normalisation, and therefore the power spectrum of the inflaton must be normalised such that
\be
{\cal P}\equiv \frac{V_N}{96\pi^2 \epsilon_N}\sim 2.46\ 10^{-9}
\ee
evaluated $\Delta N\sim 60 $ e-foldings before the end of inflation.
In the case where $\xi_\phi \Delta N \ll 1$, this reduces to
\be
{\cal P} \sim \frac{\lambda(\Delta N)^2}{72 \pi^2\xi_h ^2 }
\ee
implying that
\be
\frac{\xi_h}{\sqrt\lambda} \sim \frac{\Delta N}{6\sqrt 2 \pi \sqrt{\cal P}}.
\ee
This is of order $\xi_h \sim 4.54 \ 10^4$ for $\Delta N=60$.

\subsection{The Symmetron mechanism}
We have just seen that when $\xi_\phi \ll 1$, the end of inflation occurs close to the minimum of the effective potential. There is no dynamical reason in the previous model for the field to be there at all. Even $\Delta N$ e-foldings before the end of inflation (when $\Delta N \xi_\phi \ll 1)$ the field is still close to the minimum of the potential in the slow roll phase. One natural way of generating such a
field configuration would be to start with the field at the origin in field space when inflation begins and then evolve in the slow roll regime down to the minimum of the potential. In this case, inflation would have started
well before the required  $\Delta N\approx 60$ before its end. We now present a mechanism whereby the inflaton is stuck at the origin prior to the beginning of inflation before being released and eventually reaching the minimum of the potential.

The Higgs-dilaton inflationary model is a scalar-tensor theory with a coupling function to matter
\be
A(\varphi)= \rho^{-1} f^{-1/2} (\theta(\varphi)).
\ee
In the presence of non-relativistic matter $\rho_m$, the effective potential of the model is modified and becomes
\be
V_{\rm eff}(\varphi)=  V(\varphi) + \rho_m A(\varphi)\approx \frac{\lambda}{\xi^2_h}(1- 6\xi_\phi \theta^2) + \frac{\sqrt 6 \rho_m}{\rho}(1+ \frac{\theta^2}{2})
\ee
Let us consider the cosmological situation where the matter density dominates over the scalar potential in the mass term, i.e. $\rho_m/\rho \ge 2\sqrt 6 \lambda\xi_\phi /\xi_h^2$, implying that the curvature of the potential at the origin is positive. In this case, the effective potential
has a minimum at $\theta=0$. This forces the initial condition to be at the origin due to the matter density. Indeed,  when the matter density dominates over the scalar potential energy, the mass term of the inflaton due to the coupling to matter (in reduced units) is
\be
m^2(\rho_m)\approx  \frac{\sqrt 6 \rho_m}{\rho}
\ee
when the Hubble rate is
\be
H^2(\rho_m)\approx \frac{\rho_m}{\sqrt{6}\rho}
\ee
and much larger than the negative squared mass of the potential which is of order $\lambda\xi_\phi/\xi_h^2$.
Starting from initial values of the inflation away from the origin, the inflaton
converges to the origin in $a^{-3/2}$, i.e. very rapidly the field is stuck at the origin. Later matter stops dominating over the scalar field energy before the curvature of the potential
becomes negative at the origin (as $\xi_\phi \ll 1$).  In this case, inflation starts with a constant scalar potential energy $\frac{\lambda}{\xi_h^2}$ and an exponential growth of the scale factor which leads
the curvature at the origin from positive to negative values very quickly. From then on, slow roll inflation starts due to the a negative curvature at the origin and quantum fluctuations of the inflaton displacing its value away from zero.
This is the symmetron mechanism applied to the Higgs-Dilaton model.

In summary, as soon as  the matter density becomes subdominant, it gets redshifted in $a^{-3}$ very quickly during inflation, the field starts rolling slowly down towards the minimum of the potential and the matter density  becomes irrelevant during slow roll inflation. The symmetron mechanism implies that the inflaton starts naturally at the origin, slow rolls towards the minimum and leads to the inflationary dynamics that we have described previously when it gets close to the minimum of the potential.

\section{Conclusion}
In this paper we have introduced a new class of conformal inflation models which generalise the Higgs-Dilaton models. This class of models are scalar-tensor models. We have shown that, when non-relativistic matter is included, there is an early Damour-Polyakov mechanism which results in the inflaton field starting close to the origin of a hilltop potential. The Damour-Polyakov mechanism ensures that it is stuck here prior to the start of inflation. This automatically sets the initial conditions of inflation in this class of models. Once inflation starts the matter density is washed away. This mechanism works for a wide class of hilltop inflation models and conformal inflation models, for example it should be applicable to the superconformal inflation models \cite{Kallosh:2013lkr, Kallosh:2013hoa}.
One should note that this mechanism requires the presence of non-relativitic matter prior to inflation. In high energy models like string theory, grand unified theories or supersymmetric models with broken supersymmetry one would expect the presence of such non-relativistic matter.
The extension of our results here to such theories is currently under investigation.

We embed these models in unimodular gravity which breaks scale invariance mildly, having an effect on the late time dynamics of the model. Specialising to Higgs-Dilaton inflation models we have demonstrated explicitly the early Damour-Polyakov mechanism and analysed Higgs inflation. Inflation ends when the symmetry is broken in the Higgs potential, which is similar to the symmetron mechanism in this class of models. At late times we have shown that the dilaton field is a natural candidate for the dark energy particle with a potential due to the breaking of scale invariance in unimodular gravity. A late time  suppression of the dilaton coupling to matter implies that it  evades gravitational tests on the existence of fifth forces.

Our results should extend to other models of conformal inflation. For instance, it will be
interesting to generalise  the Damour-Polyakov mechanism to models with more than one Higgs field.
In particular one could envisage the generalisation to a supersymmetric model, such as the NMSSM\cite{Ferrara:2010yw,Ferrara:2010in}. This is under investigation.

\section{Acknowlegements}
We are grateful to Jerome Martin for discussions.
PB acknowledges partial support from the  European Union FP7  ITN
INVISIBLES (Marie Curie Actions, PITN- GA-2011- 289442) and from the Agence Nationale de la Recherche under contract ANR 2010 BLANC 0413 01. ACD is supported in part by STFC.

\bibliography{rad2}
\end{document}